\documentclass[sigconf]{acmart}

\AtBeginDocument{%
  }

\usepackage{multirow}
\usepackage{graphicx}
\usepackage{algorithm,algpseudocode}
\usepackage{makecell}
\usepackage{diagbox}
\usepackage{xcolor}
\usepackage{xspace}
\usepackage{subcaption}
\usepackage{tikz}
\usepackage{wrapfig,lipsum,booktabs}

\usepackage{custom_macros}
\usepackage{enumitem}

\newcommand{\NAME}{$\mathtt{UniGuard}$\xspace}

\setcopyright{none}
\renewcommand\footnotetextcopyrightpermission[1]{}
\begin{document}

\title{A Unified Hardware-based Threat Detector for AI Accelerators}

\author{Xiaobei Yan}
\affiliation{%
  \institution{Nanyang Technological University}
  \country{Singapore}
}
\email{xiaobei002@e.ntu.edu.sg}

\author{Han Qiu}
\affiliation{%
  \institution{Tsinghua University}
  \country{China}}
\email{qiuhan@tsinghua.edu.cn}

\author{Tianwei Zhang}
\affiliation{%
  \institution{Nanyang Technological University}
  \country{Singapore}
}
\email{tianwei.zhang@ntu.edu.sg}
\begin{abstract}
The proliferation of AI technology gives rise to a variety of security threats, which significantly compromise the confidentiality and integrity of AI models and applications. Existing software-based solutions mainly target one specific attack, and require the implementation into the models, rendering them less practical. 
We design \NAME, a novel unified and non-intrusive detection methodology to safeguard FPGA-based AI accelerators. The core idea of \NAME is to harness power side-channel information generated during model inference to spot any anomaly. We employ a Time-to-Digital Converter to capture power fluctuations and train a supervised machine learning model to identify various types of threats. Evaluations demonstrate that \NAME can achieve 94.0\% attack detection accuracy, with high generalization over unknown or adaptive attacks and robustness against varied configurations (e.g., sensor frequency and location).
\end{abstract}


\maketitle

\section{Introduction}
\label{sec:intro}
As the AI technology becomes increasingly integral to various aspects of our lives, 
its security has come under intense scrutiny. 
Researchers have discovered various security threats against AI applications, which could lead to wide-ranging consequences. For instance, by injecting adversarial perturbations (adversarial attack \cite{goodfellow2014explaining}) or poisoning training data (backdoor attack \cite{li2022backdoor}), the model will make wrong decisions. By querying the remote model with malicious samples, the adversary is able to steal the model details (model extraction attack \cite{papernot2016practical}).

Extensive studies have been conducted to combat these threats \cite{chakraborty2018adversarial,li2022backdoor,rigaki2023survey}. Existing defense solutions are mainly implemented at the software level, which suffer from several limitations. 
First, the majority of the defense strategies require to be directly integrated into the AI applications, which makes it hard to protect off-the-shelf black-box products. Second, defense at the software level is relatively less reliable, and can be subverted by many factors, such as privileged adversary, malware, memory faults, etc. Third, each defense method mainly targets one specific attack, but fails to cover other threats. Simply combining multiple defenses for different threats could incur complexity and mechanism conflict issues. These limitations underscore the need of unified, non-intrusive and reliable approaches to enhance the security of AI ecosystems. 

To this end, we introduce \NAME, a novel hardware-based methodology to detect various types of AI attacks in a holistic way. It aims to protect hardware AI accelerators, which have been widely used in many scenarios. Essentially, when an AI model is under attack, its inference behaviors can exhibit certain anomalies, giving us the potential opportunity to identify it. Prior works have observed the distinct behaviors caused by adversarial and backdoor attacks in the activation and feature space, and designed the corresponding detection tools \cite{jin2020unified,feinman2017detecting}. However, they need to collect the suspicious behaviors from inside the target model. To achieve non-intrusiveness, we posit that the attacks also leave discernible behaviors in the side-channel extraction trace, which could be monitored externally. Inspired by this, \NAME implements a Time-to-Digital Converter (TDC) as a voltage drop sensor to collect the runtime power traces of the protected model. Then it trains a machine learning model to analyze the traces and identify whether the model is being attacked, and what types of attack it is suffering. 

To our best knowledge, the only existing work that utilizes side-channel information to detect AI threats is EMShepherd \cite{ding2023emshepherd}. However, it has the following two limitations compared to \NAME: (1) EMShepherd mandates manual separation of the Electromagnetic side-channel trace for different layers in the attacked model and requires the training of a distinct classifier for each layer. In contrast, \NAME enables automatic detection with just one end-to-end model without any human intervention, this brings significant efficiency improvement. (2) EMShepherd can only detect adversarial attacks, while our \NAME is capable of identifying a diverse range of mainstream AI attacks just from one trace. 

We implement \NAME to protect the commercial-off-the-shelf (COTS) Nvidia Deep Learning Accelerator (NVDLA), and perform comprehensive evaluations. Experiment results substantiate that \NAME exhibits a remarkable detection accuracy while incurring relatively minimal overhead compared with other white-box defense methodologies. Notably, \NAME presents strong generalization in detecting unseen attacks and adaptive attacks. It also has high adaptability to different configurations.

\section{Background}
\label{sec:Background}
\subsection{Security Threats to AI Models}
The increasing ubiquity of AI systems and applications has given rise to a myriad of security challenges. Among these, three notable categories of attacks have gained prominence, as described below.  

\noindent\textbf{Adversarial attacks} \cite{goodfellow2014explaining}: This class of attacks aims to manipulate the prediction results of machine learning models by introducing subtle perturbations to input data. These perturbations are often imperceptible to human observers but can lead to incorrect model outputs. Notable algorithms employed in adversarial attacks encompass Fast Gradient Sign Method (FGSM) \cite{goodfellow2014explaining}, Projected Gradient Descent (PGD)\cite{madry2017towards}, C\&W \cite{carlini2017towards} and Deepfool \cite{moosavi2016deepfool}.

\noindent\textbf{Backdoor attacks} \cite{li2021backdoor}: These attacks involve the insertion of a backdoor into a machine learning model. This can be typically realized by training data poisoning. Such backdoor remains dormant under normal circumstances but can be activated with any input samples containing a pre-defined trigger. Then the model will make wrong predictions as desired by the attacker. 

\noindent\textbf{Model extraction attacks} \cite{papernot2016practical}: These attacks focus on stealing the proprietary or sensitive information (e.g., network structure, hyper-parameters, parameters) from a machine learning model. This is achieved by sending special samples to query the victim model, and extract the information from the returned responses. 
The query samples used for model extraction fall into different categories \cite{kariyappa2020defending}.
(1) \textit{Synthetic data}: the adversary synthesizes data from a small set of in-distribution seed samples. These synthetic data samples are produced iteratively, incorporating heuristic-based perturbations, such as Jacobian-Based Data Augmentation (JBDA) \cite{papernot2016practical} or Random Synthetic Sample Generation, into the seed examples.
(2) \textit{Surrogate Data}: the adversary employs a surrogate dataset to query the target model. For instance, a cat dataset can be employed as a surrogate dataset to query a dog-breed classifier.

\subsection{Nvidia Deep Learning Accelerator (NVDLA)}
NVDLA is a versatile open-source architecture developed by Nvidia for enhancing deep learning inference. It boasts the capability to perform various operations in model inference, e.g., convolution, activation, pooling, normalization. The adaptability of NVDLA is evident in its configurability, allowing for both large and small implementations. These configurations differ in the core dimensions and implementation of specific engines, e.g., Rubik and DMA.

The architecture of NVDLA  is delineated into two fundamental components: hardware design and software design. The hardware design comprises a series of pipeline stages housing diverse types of engines that govern the behavior of FPGA boards. The software design acts as an intermediary between users and hardware components. Its primary responsibility is to construct and load the AI model onto the FPGA board for execution.

\subsection{Power Side Channel on FPGA}
Malicious actors frequently employ power side-channel analysis as a non-intrusive reverse engineering method to compromise the security of cryptographic systems. Besides, this technique can also serve legitimate purposes, allowing security experts to assess the effectiveness of hardware security mechanisms, and ensuring that sensitive data remain confidential and resistant to power attacks.

Power analysis is also prevalent for FPGA devices. The underlying principle is that the power consumption of an FPGA chip varies based on the specific operations being executed. These fluctuations may inadvertently leak information pertaining to internal operations, data, and algorithms. Specifically, most components on an FPGA chip share a common Power Distribution Network (PDN). This PDN can be represented as an RLC circuit, where a resistor (R), an inductor (L), and a capacitor (C) are connected either in series or parallel. The intensive switching activities on the chip can lead to voltage fluctuations within the PDN. The transient voltage drop experienced by a circuit can be modeled as $V_{drop}=IR+L\frac{di}{dt}$, where $L\frac{di}{dt}$ reflects the impact of switching activities on the FPGA \cite{zhao2018fpga}.
Typically, in CMOS circuits, the logical delay of combinational logic gates is inversely proportional to the voltage supplied to each gate, based on which we can infer the switching activities. 

\NAME utilizes a Time-to-Digital Converter (TDC) to measure the combinational logic delay. The TDC employs a clock signal that propagates through a chain of buffers, serving as the voltage drop sensor. Discrepancies in switching activities for various calculations in different parts of the FPGA lead to variations in voltage drop values, resulting in different delay measurements in the TDC. These distinct delays influence the propagation lengths within the delay line, which affect the values in the latches. Consequently, the activities of other circuits on the FPGA can be identified through the TDC readout, as demonstrated in prior studies \cite{gravellier2021remote}.

\section{\NAME}\label{sec:details}
\label{sec:detector}
As a novel threat detector, \NAME is designed to satisfy the following requirements. 

\begin{itemize}[leftmargin=*]
\item\textbf{Unified:} \NAME serves as a universal detector capable of identifying multiple threats to AI models, significantly reducing the cost of attack prevention. 

\item\textbf{Non-intrusive:} \NAME is a hardware-based solution. It treats the protected model as a black box. It does not require any modifications and implementations inside the model. It only needs to set up a TDC voltage sensor on the same FPGA board as the accelerator, which only passively collects the power trace without interfering the model execution. 

\item\textbf{Platform-agnostic:} \NAME is agnostic to the FPGA board, DNN accelerator implementation, the model and task. Its hardware design is an IP block that can be seamlessly integrated into the target platform, while its software design operates as an independent driver, separate from the accelerator's software. This plug-and-play (PnP) feature facilitates easy portability to a wide range of hardware devices and applications.

\item\textbf{Automatic:} \NAME can automatically detect the attacks in real-time, without any user intervention. This is different from EMShepherd \cite{ding2023emshepherd}, which requires extensive manual preprocessing of Electromagnetic traces.

\item\textbf{Robust:} \NAME demonstrates robustness against varied platform configurations. 

\item\textbf{Generalizable:} \NAME is effective in detecting any unseen attacks and adaptive attacks. 
\end{itemize}

\subsection{Overview}
\setlength{\columnsep}{7pt}%

\NAME comprises two key phases, as shown in Figure \ref{framework}. In the training phase, the defender utilizes a public dataset to simulate the attack they aim to detect and collects the power side-channel traces to train the detector. Subsequently, the trained detection model is deployed in real-world scenarios to detect potential attacks with a single power trace obtained from the sensor.

\NAME operates in two distinct phases. In the \textit{profiling} phase, we initiate a series of randomized model generation processes using a public dataset. We set up a TDC on the same FPGA board to collect the power traces of normal inference executions from these models. Then we launch various attacks against these generated models on the AI accelerator, and use the TDC to collect the corresponding malicious power traces. The normal and malicious power traces form a dataset, from which we train the detection model. 

In the \textit{detection} phase, we use the TDC to capture the power trace of each inference process, and feed the trace to the detection model. This detection model can determine whether the victim model is currently under any attack. 

\subsection{Power Monitor Module} 
\label{sec:detector.tdc}
In \NAME, a Power Monitor Module is required to collect the inference execution traces in the profiling phase to build and detection model, and capture the real-time execution trace of the victim model in the detection phase. Following \cite{yan2023mercury}, we employ a Time-to-Digital Converter (TDC) as the power sensor.

\begin{figure}
\centering
\begin{minipage}{.48\linewidth}
  \centering
  \includegraphics[width=\linewidth]{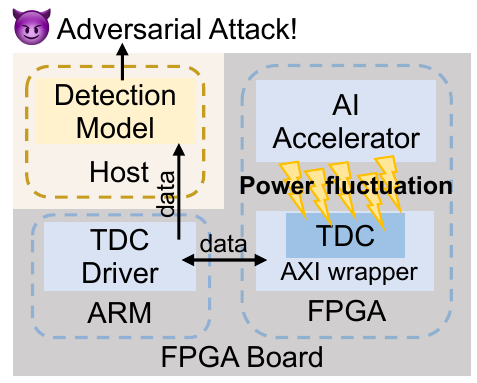}
  \vspace{-20pt}
  \captionof{figure}{\NAME Overview}
  \label{framework}
\end{minipage}%
\hspace{1pt}
\begin{minipage}{.48\linewidth}
  \centering
  \vspace{10pt}
  \includegraphics[width=\linewidth]{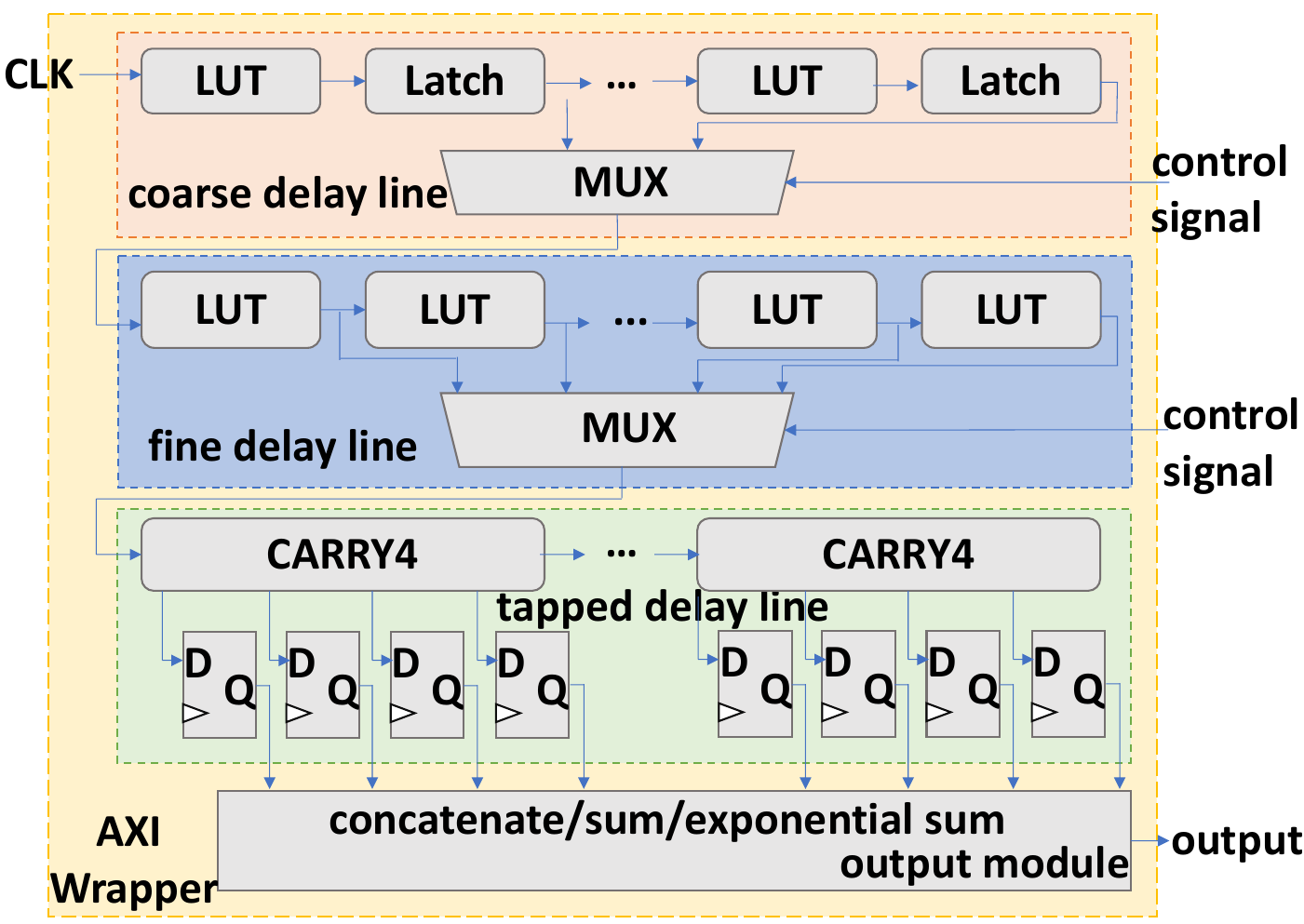}
  \vspace{-18pt}
  \captionof{figure}{TDC architecture}
  \label{tdc-detail}
\end{minipage}
\vspace{-15pt}
\end{figure}

Figure \ref{tdc-detail} offers an intricate insight into the TDC architecture. In this design, the incoming clock signal traverses an adjustable coarse delay line and a fine delay line. These elements collectively contribute to establishing an initial delay, which then is fed into a tapped delay line. The adaptability of the initial delay is achieved through dynamic configuration, facilitated by multiplexers (MUX). The calibration process involves modifying the number of logic elements constituting the coarse and fine delay lines, enabling customization of the delay duration.

The coarse delay line, comprising replicated Look-Up Table (LUT) and latch modules, offers a substantial delay. The fine delay line, equipped with replicated LUT modules, provides a finer degree of control over the delay. The tapped delay line employs carry chains and leverages \verb!CARRY4! primitives, with their \verb!CO! outputs registered by four dedicated D flip-flops. During each readout, this component monitors the taps reached by the clock signal, yielding a raw value. Depending on the configuration specified in the TDC IP settings, this raw output can be concatenated or transformed into a sum or exponential sum.

It is critical to perform the TDC calibration, particularly the adjustment of its initial delay, which precedes the output measurements. Our calibration process, embedded within the TDC driver, operates in two loops. It systematically explores all conceivable combinations of fine and coarse delay line lengths, determining the optimal initial delay value. This ensures that the signal remains within the delay line when its state is captured by the register.

\subsection{Detection Model}
\label{sec:detector.model}

\noindent\textbf{Power Trace Pre-processing.} 
Before model training or inference, we need to first preprocess the collected trace with two essential operations: averaging the data and reshaping them from a one-dimensional array into a two-dimensional format. 

Specifically, the raw side-channel data have high-frequency fluctuations. Averaging helps reduce the noise or small-scale variations in the raw data, and then enhances the model's ability to capture broader patterns and features. 
Then, we normalize the data and organize them into three rows. The conversion to a matrix structure renders it more suitable for processing by convolutional and recurrent layers, facilitating subsequent model operations. 

\begin{figure}[t]
	\centering
	\includegraphics[scale=0.36]{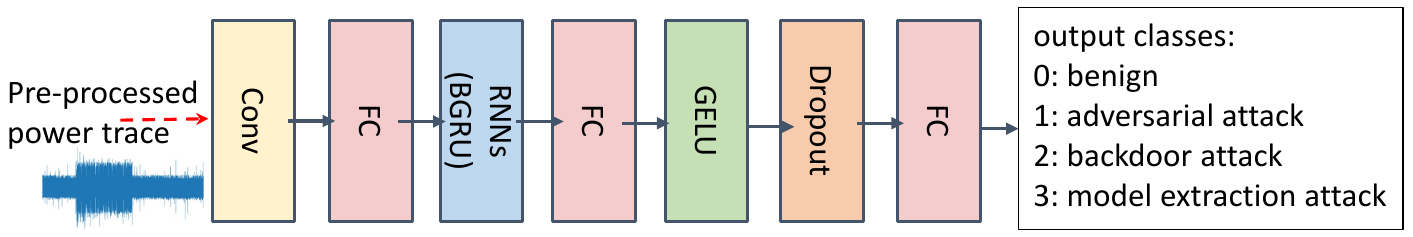}
    \vspace{-15pt}
 \caption{Detection model architecture in \NAME.}
    \vspace{-10pt}
	\label{model}
\end{figure}

\noindent\textbf{Detection Model Architecture.} 
Figure \ref{model} shows the detailed network architecture of \NAME. 
The detection model can effectively analyze power side-channel traces and classify them into distinct categories: (0) benign, (1) adversarial attack, (2) backdoor attack, (3) model extraction attack. 

The preprocessed power trace is first fed into a convolution layer, which is responsible for extracting essential features. 
Following this, a fully connected layer is employed to transform the extracted features into a format suitable for further processing. 
The model incorporates multiple Recurrent Neural Networks (RNNs) with Bidirectional Gated Recurrent Unit (BGRU) cells. These RNNs are well-suited for capturing temporal dependencies and sequential patterns in the traces. The bidirectional nature of the GRU cells enables the model to consider both past and future contexts, enhancing its ability to discern subtle differences.
The Gaussian Error Linear Unit (GELU) activation function is applied to model the complex non-linear relationships within the data.
To mitigate overfitting and enhance model generalization, a dropout layer is integrated. This operation randomly deactivates a fraction of neurons during training, forcing the model to rely on different pathways and reducing its susceptibility to side-channel noise.
The final fully connected layer serves as the output layer, where the model assigns one of the four class labels to the input sample. 

\noindent\textbf{Interpreting the Detection Mechanism.}
We further leverage Class Activation Map (CAM) to interpret the understand why our detection model can distinguish different types of traces. It sheds light on the regions of the power trace that significantly influence the model's prediction. Specifically, we employ the Grad-CAM \cite{selvaraju2016grad} technique to visualize and interpret the activations of the detection model. This algorithm involves the selection of a target layer, attaching hooks to that layer for both forward and backward passes during inference, gradient calculation, and weighted summation.

\begin{figure}[t]
  \centering
  \begin{subfigure}{.5\linewidth}
    \includegraphics[width=\linewidth]{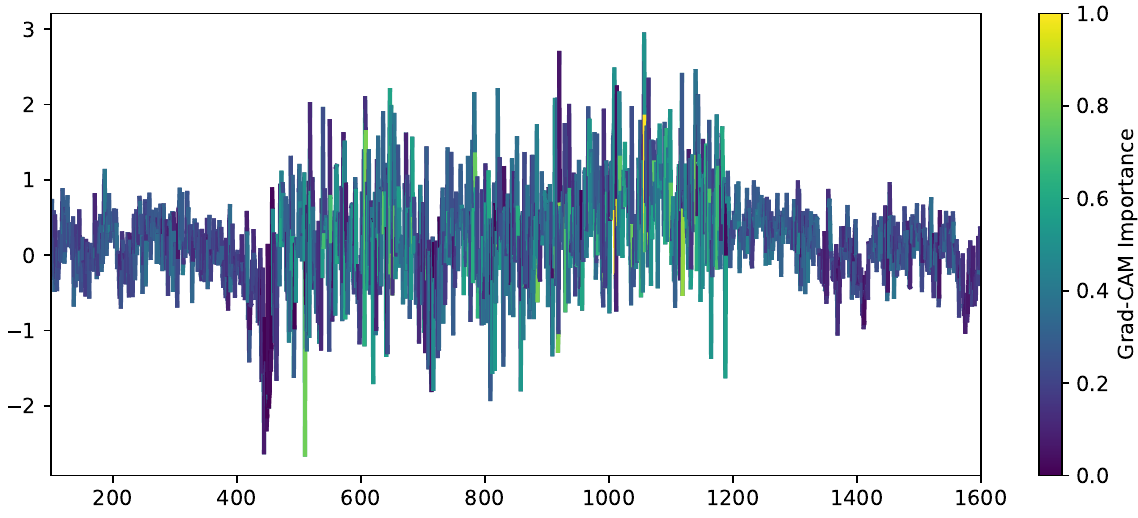}
    \caption{benign input}
    \label{fig:sub1}
  \end{subfigure}\hfill
  \begin{subfigure}{.5\linewidth}
    \includegraphics[width=\linewidth]{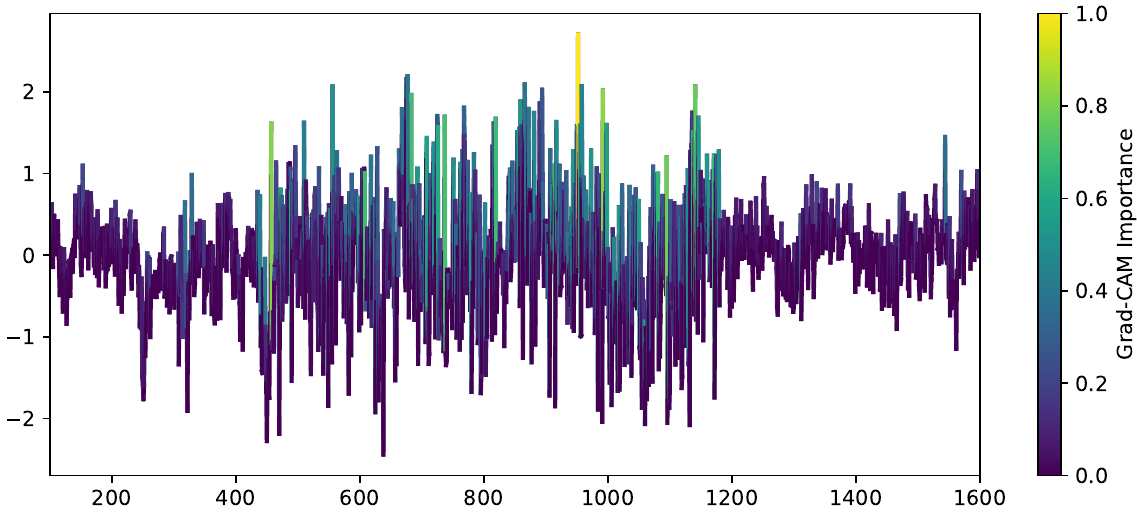}
    \caption{adversarial attack}
    \label{fig:sub2}
  \end{subfigure}

  \medskip
  \vspace{-5px}

  \begin{subfigure}{.5\linewidth}
    \includegraphics[width=\linewidth]{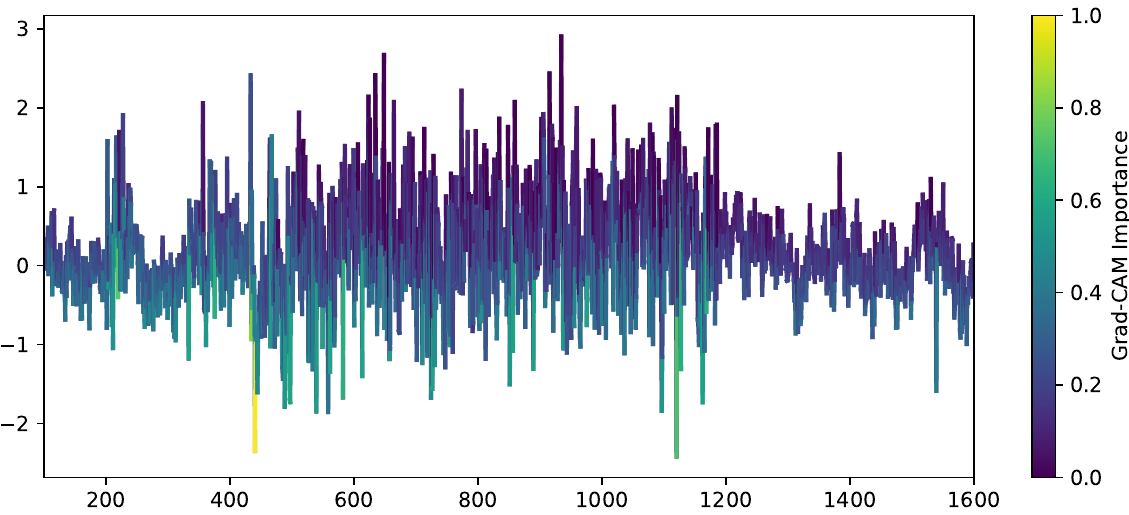}
    \caption{backdoor attack}
    \label{fig:sub3}
  \end{subfigure}\hfill
  \begin{subfigure}{.5\linewidth}
    \includegraphics[width=\linewidth]{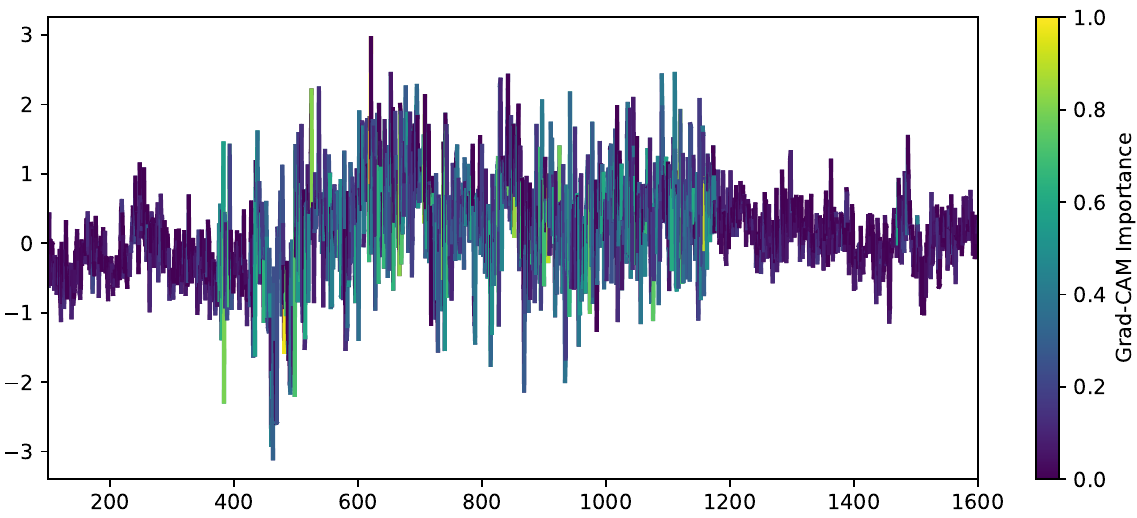}
    \caption{model extraction attack}
    \label{fig:sub4}
  \end{subfigure}
  \vspace{-20pt}
  \caption{CAM for 4 output classes.}
  \vspace{-15pt}
  \label{cam}
\end{figure}

Figure \ref{cam} presents the CAMs for four output classes. The curve in each figure represents the input trace to the model. Different colors along the curve signify the varying degrees of importance, with the brightest color indicating the most influential regions. The results illustrate that \NAME concentrates on the central part, aligning with the trace segment where FPGA calculations are ongoing. Notably, the significant regions appear discretely throughout the trace, indicating that \NAME's decision is informed by a combination of information from all layers—a characteristic akin to software-based detection methods. This also makes it more difficult to conduct adaptive attacks to bypass the detection (Section \ref{sec:adaptive-attack}). 

\section{Evaluation}\label{sec:evaluation}
\noindent\textbf{Testbed.} 
In our experiments, we choose the Xilinx Zynq-7000 SoC ZC706 board (xc7z045ffg900-2) as our testbed with the small NVDLA implementation. The board runs Ubuntu 16.04 OS, and Vivado 2019.1 is used for hardware design. NVDLA operates at a clock frequency of 10MHz, and the TDC sensor clock is set to 150MHz, while the TDC AXI clock runs at 10MHz. Data exchange between the board and the host computer is facilitated through Ethernet using the \verb!scp! command. For model training and execution, we utilize PyTorch version 1.13 and CUDA version 11.6, running on a server equipped with a Nvidia GeForce RTX 3090 GPU.

\noindent\textbf{Detection Dataset Construction.} 
We create a dataset with normal and various types of attack traces for training and testing the detection model. Initially, we randomly generated 400 models on the MNIST dataset and deployed them on the NVDLA accelerator. These models have random numbers (in the range of [2, 18]) and types of network layers, featuring 12 different convolution layers with varying kernel sizes (2, 3, 4, 5) and output sizes (10, 20, 30), 4 pooling layers with different kernel sizes (2, 3, 4, 5), 5 fully-connected layers with varying output sizes (100, 200, 300, 400, 500), 1 ReLU layer, and 1 softmax layer. These models are first pre-trained using Caffe, then calibrated by TensorRT, and finally compiled by the NVDLA compiler. They are generated on the host computer and executed by NVDLA runtime on the FPGA. 

Targeting these models, we launch three attacks: adversarial, backdoor and model extraction attacks. We capture the corresponding traces together with normal inference to construct a dataset. For each type of attack, we choose three state-of-the-art methods. Table \ref{attack_params} lists these methods and hyper-parameters, such as perturbation magnitude ($\epsilon$), norm ($L_2$), step size ($\alpha_s$), watermark strength ($\alpha$), constant ($c$), and objective function ($f$). 
We partition each class of traces into two parts: 90\% for training and 10\% for testing. 

\begin{table}[htbp]
  \centering
  \vspace{-10pt}
  \caption{Attack parameters}
  \vspace{-10pt}
   \resizebox{\linewidth}{!}{%
    \begin{tabular}{lll}
    \toprule
    \textbf{Attack} & \textbf{Method} & \textbf{Hyper-parameters} \\
    \midrule
          & FGSM  & $\epsilon=0.5$ \\
    Adversarial & PGD   & $\epsilon=0.5, L_2, \alpha_s=8/255$ \\
    Attack & C\&W & $L_2, c=0.01-10^{10},f=f_6$ \\
    \midrule
            & Pattern trigger   & $\alpha=0.4$, poison rate=10\% \\
    Backdoor& Instance trigger     & poison rate=1.7\% \\
    Attack  & Watermark & $\alpha=0.4$, poison rate=10\% \\

    \midrule
    Model & Surrogate (FashionMNIST) & 28*28 grayscale image \\
    Extraction  & Surrogate (CIFAR-10) & 28*28 grayscale image\\
    Attack   & Synthetic (JBDA)  & $\lambda=0.1,lr=5*10^{-3},$ epoch=10 \\
    \bottomrule
    \end{tabular}%
    }
    \vspace{-15pt}
  \label{attack_params}%
\end{table}%
 
\subsection{Detection Accuracy}

We first evaluate the impact of hyperparameters on the performance of the detection model in \NAME. We primarily focus on two key parameters: number of RNN layers $N$ (ranging from 1 to 6) and RNN dimension $D$ (128 or 256). We train the model with each configuration over the constructed dataset for 100 epochs. The results are presented in Table \ref{RNN-hyperparameter}. We observe that more RNN layers can significantly improve the detection accuracy, with 5 RNN layers achieving the best results. Moreover, in most cases, an RNN dimension of 128 outperforms that of 256. Therefore, we adopt the configuration of 5 RNN layers and a dimension of 128 for the detection model in the following experiments.

\begin{table}[htbp]
\begin{subtable}[c]{0.48\linewidth}
\centering
    \resizebox{0.8\linewidth}{!}{
    \begin{tabular}{rrr}
    \toprule
    \multicolumn{1}{l}{\textit{N}} & \multicolumn{1}{l}{Train Acc} & \multicolumn{1}{l}{Test Acc} \\
    \midrule
    1     & 92.8  & 67.8 \\
    2     & 97.0  & 81.4 \\
    3     & 98.1  & 86.6 \\
    4     & 98.9  & 88.6 \\
    5     & 99.3  & \textbf{91.0} \\
    6     & 99.3  & 89.7 \\
    \bottomrule
\end{tabular}}
\subcaption{Results for \textit{D}=128}
\label{hyperparameters1}
\end{subtable}
\begin{subtable}[c]{0.48\linewidth}
\centering
    \resizebox{0.8\linewidth}{!}{
    \begin{tabular}{rrr}
    \toprule
    \multicolumn{1}{l}{\textit{N}} & \multicolumn{1}{l}{Train Acc} & \multicolumn{1}{l}{Test Acc} \\
    \midrule
    1     & 98.8  & 66.6 \\
    2     & 98.2  & 81.2 \\
    3     & 98.7  & 85.7 \\
    4     & 98.9  & 85.8 \\
    5     & \textbf{99.6 } & 89.0 \\
    6     &  99.1     & 89.8 \\
  \bottomrule
\end{tabular}}
\subcaption{Results for \textit{D}=256}
\label{hyperparameters2}
\end{subtable}
\vspace{-5pt}
\caption{Impact of RNN configurations}
\label{RNN-hyperparameter}
\vspace{-25pt}
\end{table}

Second, we compare \NAME with state-of-the-art AI detection methods. These include adversarial attack detection schemes: HASI \cite{samavatian2021hasi}, EMShepherd \cite{ding2023emshepherd}, Feature Squeezing (FS) \cite{xu2017feature}, Kernel Density Estimation (KDE) \cite{feinman2017detecting}, and \cite{odetola2022hardening}, along with backdoor attack detection schemes \cite{kwon2020detecting,fu2020detecting,xu2021detecting}. We highlight that such comparisons are not quite fair for \NAME, as these baseline methods have more requirements or limitations: (1) they require multiple inference queries to detect one attack, while \NAME only needs to analyze one query; (2) some of the methods (e.g., KDE, FS, HASI and \cite{fu2020detecting,odetola2022hardening}) require extra information about the target model, including the intermediate outputs, testing inputs, or complete knowledge of the model architecture and parameters. \NAME does not need to have such information; (3) these methods are designed to detect \textit{one} specific type of attack, while \NAME is able to cover all. 

Table \ref{main_acc} shows the comparison results. Note that some model extraction attacks also employ adversarial attack methods to synthesize query samples \cite{juuti2019prada}, making them fundamentally indistinguishable. Therefore, we also report the combined accuracy of adversarial and model extraction attacks (the \NAME* row). We observe that although \NAME has fewer requirements than existing methods, it still has superior detection accuracy. In the following evaluation, we will mainly focus on the results of \NAME*.

\begin{table}[htbp]
  \centering
  \vspace{-10pt}
  \caption{Detection Accuracy}
  \vspace{-10pt}
  \resizebox{\linewidth}{!}{
    \begin{tabular}{llllll}
    \toprule
    \multicolumn{1}{c}{\multirow{2}[4]{*}{Method}} & \multicolumn{4}{c}{Detection Accuracy(\%)} & Total \\
\cmidrule{2-5}          & Benign & Adversarial & Backdoor & Model extraction & Acc \\
    \midrule
    HASI  &   -    & 87 &  -     & -      &  -\\
    EMShepherd &   -    & \textbf{94} &     -  &  -     &  -\\
    FS    &   -    & 67 &   -    &    -   &-  \\
    KDE   &    -   & 50 &    -   & -      &  -\\
    \cite{odetola2022hardening}   &    -   & 71 &    -   & -      &  -\\
    \cite{kwon2020detecting} & 79 &  -     & 81 & -      &-  \\
    \cite{fu2020detecting} &  -     &   -    & 90 & -      & -  \\
    \cite{xu2021detecting} &  -     &   -    & 90 & -      &  - \\

    \NAME & 97.4 & 68.6 & \textbf{94.1}& 92.0 & 87.9 \\
    \NAME* & 97.4 & 92.1 & \textbf{94.1} & 92.1 & 94.0 \\
    \midrule
    \multicolumn{6}{l}{False Positive Rate (FPR): EMShepherd, KDE=10\%, HASI=6\%, FS=4.5\%} \\
    \multicolumn{6}{l}{*: Combined accuracy of adversarial and model extraction attack} \\
    \end{tabular}}%
      \vspace{-20pt}
  \label{main_acc}%
\end{table}%

\subsection{Resource Overhead}
We evaluate the resource overhead incurred by \NAME. Given that there are very few hardware-based detection methods, we compare our design with the adversarial attack detector in \cite{odetola2022hardening}. The results are summarized in Table \ref{util}, where "Area Consumed" represents the ratio of Look-Up Tables (LUT) and flip-flops (FF) to the corresponding available resources on the FPGA. It is obvious that \NAME imposes much lower hardware resource overhead.
\begin{table}[htbp]
  \centering
  \vspace{-10pt}
  \caption{Overhead comparisons}
  \vspace{-10pt}
  \resizebox{\linewidth}{!}{
    \begin{tabular}{lrrrll}
    \toprule
    Solution & \multicolumn{1}{l}{LUT} & \multicolumn{1}{l}{FF} & \multicolumn{1}{l}{BRAM} & \multicolumn{1}{l}{DSP}  & Area Consumed (\%) \\
    \midrule
    \cite{odetola2022hardening} & 17510  & 8528  & 2001     & 40    & 32.9\% / 8.01\% \\

    \NAME  & 1051  & 1505   & 0     & 0    & 0.48\% / 0.34\% \\
    \bottomrule
    \end{tabular}}%
  \vspace{-10pt}
  \label{util}%
\end{table}%

\subsection{Robustness}\label{sec:evaluation.robustness}
We assess \NAME's robustness under different configurations. 

\noindent\textbf{Clock Frequency.}
We first reduce the working frequency of the AXI bus for the power monitor sensor and investigate the impact on the detection accuracy. Lowering the frequency setting for the AXI bus results in a reduced amount of data being collected. As explained in Section \ref{sec:detector.model}, we perform an averaging operation to preprocess the data. So we select two window sizes for averaging (50 and 10), and the detection results are shown in Table \ref{freq_result}. The "factor" column represents the ratio of the original frequency to the experimental frequency. The results indicate that \NAME maintains its effectiveness even with a lower frequency for the AXI bus of the power monitor. This underscores the robustness and adaptability of \NAME across varying operational frequencies.

\begin{table}[htbp]
\vspace{-10pt}
\begin{subtable}[c]{0.48\linewidth}
\centering
    \resizebox{\linewidth}{!}{
    \begin{tabular}{rrr}
    \toprule
    \multicolumn{1}{l}{Factor} & \multicolumn{1}{l}{Train Acc} & \multicolumn{1}{l}{Test Acc} \\
    \midrule
    2     & 98.7 & 83.9 \\
    3     & 93.0  & 54.1 \\
    4     & 97.9  & 76.4 \\
    5     & 98.6  & 83.4 \\
    \bottomrule
\end{tabular}}
\subcaption{Averaging Window: 50}
\label{freq_result_50}
\end{subtable}
\begin{subtable}[c]{0.48\linewidth}
\centering
    \resizebox{\linewidth}{!}{
    \begin{tabular}{rrr}
    \toprule
    \multicolumn{1}{l}{Factor} & \multicolumn{1}{l}{Train Acc} & \multicolumn{1}{l}{Test Acc} \\
    \midrule
    2     & 98.0 & \textbf{93.2} \\
    3     & 99.4  & 87.9 \\
    4     & \textbf{99.5}  & 86.8 \\
    5     & 99.2  & 89.2 \\
  \bottomrule
\end{tabular}}
\subcaption{Averaging Window: 10}
\label{freq_result_10}
\end{subtable}
\vspace{-5pt}
\caption{Impact of RNN configurations}
\label{freq_result}
\vspace{-20pt}
\end{table}

\noindent\textbf{TDC Placement Location.}
Next, we investigate the impact of TDC locations on the FPGA board during model extraction. Prior studies \cite{0Understanding, gravellier2021remote} have indicated the sensitivity of TDC outputs to its placement. It is crucial to identify the optimal location for deploying the TDC. We explore three different TDC locations on FPGA: top-left, center, and bottom-right. We use Pblock in Vivado to set the location constraints. Table \ref{tdcloc} shows the detection accuracy with different TDC locations ("Acc. (w/o. Aug)" column). We observe that the TDC location can indeed affect the detection accuracy, given the variations in side-channel power traces. 

A potential solution to mitigate such impact is to augment the training dataset with power traces collected from multiple locations. The trained model will be more general and robust against the actual TDC placement at real time. Table \ref{tdcloc} ("Acc. (w/. Aug)" column) shows the enhanced results where we augment the dataset with 10\% traces for each of the three locations. It is obvious that dataset augmentation gives a significant improvement in the detection performance, approaching to that of the original location.

\begin{table}[htbp]
  \centering
  \vspace{-5pt}
  \caption{Sensitivity to TDC locations}
  \vspace{-10pt}
    \begin{tabular}{lrr}
    \toprule
    TDC Locations & \multicolumn{1}{l}{Acc. (w/o. Aug)} & \multicolumn{1}{l}{Acc. (w/. Aug)} \\
    \midrule
    top-left &  41.0\%     & 70.1\% \\
    center &   46.0\%    &  74.7\%\\
    bottom-right &  48.2\%     & 91.2\% \\
    \bottomrule
    \end{tabular}%
    \vspace{-10pt}
  \label{tdcloc}%
\end{table}%

\section{Generalization to More Attacks}
\label{sec:adaptive-attack}

\subsection{Unseen Attacks}
When training the detection model, we collect the malicious traces of different attack methods to construct the training dataset. It is important that the detection model is capable of detecting other attacks not included in the training as well. To test the generalization of \NAME, we measure its detection accuracy against three unencountered attacks. Specifically, for adversarial attack, we choose the Deepfool method; for backdoor attack, we choose a square of 3*3 pixels as a new trigger design; for model extraction attack, we choose the CIFAR-100 as the surrogate data. 
Our experiments reveal that \NAME can achieve the detection accuracy of 95.6\% for benign samples, 62.6\% for the new adversarial and model extraction attacks, and 83.1\% for the new backdoor attack. This reveals that \NAME can effectively generalize to new and unanticipated attack methods.

\subsection{Adaptive Attacks}

We consider a more sophisticated scenario, where a smart attacker knows the mechanism of our defense (not the detection model parameters) and tries to bypass the detection. We investigate whether \NAME is able to detect such attacks as well. 

To achieve this, we follow \cite{jain2022adversarial} to craft the Detection Avoidance Attack against \NAME. Formally, let $D$ be the detection model, $f$ be the target victim model, and $X$ be a malicious input for one attack. Then a good detection model satisfies $D(Tr(X)) = \text{attack}$, where $Tr(X)$ is the corresponding power trace of $X$. In order to bypass the detection, the attacker needs to find a small perturbation $\delta$ and add it to $X$, which makes $D(Tr(X)) = \text{benign}$. Meanwhile, the new input should still keep the same attack effects: $f(X+\delta) = f(X)$. The attacker also aims to make the scale of $\delta$ as small as possible so the new input $X+\delta$ still keeps similar semantics as $X$. 

Since $f$ and mapping between the input and power trace is unknown to the attacker, he cannot directly identify the optimal $\delta$. Instead, he can leverage state-of-the-art black-box adversarial attack techniques. Algorithm \ref{algo1} shows the detailed optimization step, which consists of two phases.

Formally, we aim to find $\delta$ that satisfies the following optimization problem:
\begin{equation}
\begin{split}
\text{find }& \delta, \text{ s.t. } \\
L(D(\text{Tr}(X+ \delta), & \text{class}_{\text{benign}})) \rightarrow \min \text{, }\\
f(X+ \delta) &= f(X)
\end{split}
\end{equation}

Here, $f$ denotes the victim's model running in the accelerator, $\text{Tr}(X')$ is the power side-channel trace of input $X'$, $D$ is the detector's model, and $L$ is the loss function of the detector. In our specific case, the goal is to make the detector output the benign class, necessitating a targeted adversarial attack with gradient descent. 

\begin{algorithm}[h]
\footnotesize
\caption{Detection avoidance attack at the ${t}^{th}$ iteration}
\label{algo1}
    \begin{algorithmic}[1]
        \Inputs{
        $L$: Loss function of detection model to the target (benign) class. \\
        $X$: Original input to be attacked, of size $n$. \\
        $\delta_{t}$: Point at which the gradient is to be estimated. \\
        $d'$: Number of Gaussian samples (should be even). \\
        $\sigma$: Scaling factor for Gaussian samples $\sim N(0, I_n)$.\\
        $grad_{t-1}$: Weighted sum of previous gradients.\\
        $\epsilon$: Upper bound on $||\delta||_p$.\\
        $\mu$: Momentum parameter.\\
        $\eta$: Step size to update $\delta$ in each iteration.
        }
        
        \Output{
        $grad_{t}$: $\nabla_{\delta}{E[L(\delta)]}$, estimate of the gradient of $L(\delta)$\\
        $\delta_{t+1}$: Perturbation after the ${t}^{th}$ iteration.\\
        }
    
        \Initialize{
        $\theta_i \gets N(0,I_n)$, for $i \in \{1, ...,\frac{d'}{2}\}$\\
        $\theta_i \gets -\theta_{d'-i+1}$, for $i \in \{(\frac{d'}{2} + 1), ...,d'\}$\\
        $grad_t \gets 0$
        }
    
        \For{$i = 1$ to $d'$}
            \State{
            $\theta_{i}' \gets \max ( \min (1, X + \delta_t + \sigma \theta), 0)  - X - \delta_t $}
            \State{
            $grad_t \gets grad_t + L(X+\delta_t + \theta_i') * \theta_i'* \frac{1}{{\sigma}d'}$
            }
        \EndFor
        \State{$grad_t \gets \mu * grad_{t-1} + (1 - \mu) * grad_t$}
        \State{$\delta_{t+1} \gets \delta_t - \eta * \textrm{sign}(grad_t)$}
        \State{$\delta_{t+1} \gets \min(\max(X + \delta_{t+1}, 0), 1) - X$}
        \If {$||\delta_{t+1}||_p > \epsilon$}
        \State{$\delta_{t+1} \gets \delta_{t+1} * \epsilon / ||\delta_{t+1}||_p$}
        \EndIf
        \State{$grad_{t-1} \gets grad_t$}\\
        \Return $grad_t, \delta_{t+1}$
    \end{algorithmic}
\end{algorithm}

(1) \textit{Gradient Estimation.}
The attacker performs zero-order gradient estimation through Natural Evolutionary Strategies (NES) \cite{salimans2017evolution}, which can be reviewed as a specific instance of finite-differences estimation on a random Gaussian basis. This is commonly used for optimization under the black-box settings \cite{ilyas2018black}. Let $L$ be the loss function of the detection model. Then the gradient of $L$ can be estimated using the following equation:
\begin{equation}\label{equation:gradient_estimate}
\nabla_{\delta}{E[L(\delta)]} \approx \frac{1}{\sigma d}\sum_{i=1}^{d'}\theta_iL(\delta + \sigma\theta_i)
\end{equation}
Here, $\theta_i\sim N(0, I_n), 1 \leq i \leq d'$ represents samples drawn from a standard multivariate normal distribution over $\mathbb{R}^n$. To reduce the variance in our estimation, the attacker employs antithetic sampling by generating Gaussian noise samples $\theta_i$ for $i \in {1,...,\frac{d'}{2}}$ and setting $\theta_j = -\theta_{d'-j+1}$ for $j \in {(\frac{d'}{2} + 1),...,d'}$, where $d'$ is an even number (line 3). These samples are utilized to query the target model and obtain the power traces. Subsequently, the attacker feeds these traces into the detection model to calculate the loss $L(\delta + \theta_i')$ and estimate the gradient at this specific point (lines 5-6).

(2) \textit{Perturbation Update.}
The attacker updates $\delta_{t}$ at each step $t$, using the sign of the estimated gradient $sign(grad)$ with a momentum parameter $\mu$ (lines 8-9). Clipping is applied to ensure the resulting input $X + \delta_{t+1}$ remains within the boundary (line 10). 

\begin{figure}[t]
	\centering
    \includegraphics[scale=0.4]{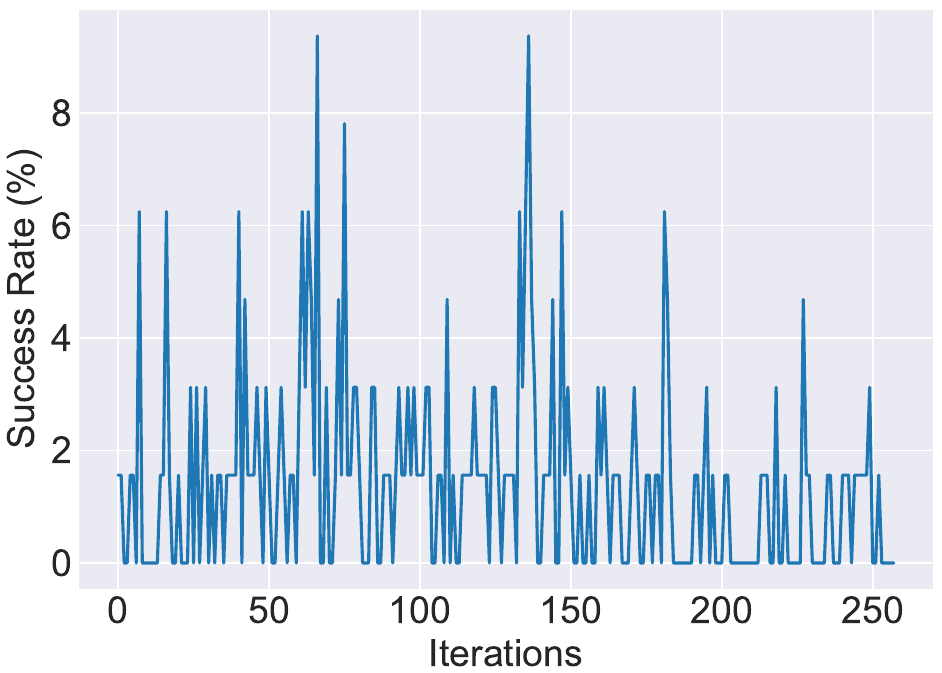}
    \vspace{-10pt}
    \caption{Attack success rate of be predicted as benign.}
	\label{Detection_Avoidance_Attack_result}
    \vspace{-20pt}
\end{figure}

\noindent\textbf{Evaluation results.}
We implement such attack with $d'=256$ samples generated at each iteration of Algorithm \ref{algo1}, comprising a total of $t=256$ iterations, resulting in a maximum query budget of 65,536. We set the scale of Gaussian noise $\sigma=0.001$, learning rate $\eta=0.001$, momentum term $\mu=0.5$, and $\epsilon=1/255$. For each step \textit{t}, we repeatedly collect the power trace of $X + \delta_{t}$ for 100 times, and compute the average accuracy of being detected as benign. The results are shown in Figure \ref{Detection_Avoidance_Attack_result}. It is obvious that the accuracy is close to 0 towards the benign class as the query budget reaches 65,536, indicating the ineffectiveness of such attack against \NAME. 

Several factors contribute to \NAME's resilience against adaptive attacks. Notably, the inherent noise within the power measurements introduces complexity and unpredictability in estimating the gradients, making it hard to identify qualified perturbations. 

\section{Conclusion}
This paper presented \NAME, a novel hardware-based methodology to protect AI accelerators. \NAME exhibits the capability of detecting a spectrum of AI attacks, utilizing side-channel information captured by a TDC during model inference. It is non-intrusive to the target AI application, and easy to use and deploy. Experiments demonstrate \NAME achieves high detection accuracy, robustness and generalization to various attacks.

\bibliographystyle{ACM-Reference-Format}
\bibliography{mybib}


\begin{thebibliography}{28}


\ifx \showCODEN    \undefined \def \showCODEN     #1{\unskip}     \fi
\ifx \showDOI      \undefined \def \showDOI       #1{#1}\fi
\ifx \showISBNx    \undefined \def \showISBNx     #1{\unskip}     \fi
\ifx \showISBNxiii \undefined \def \showISBNxiii  #1{\unskip}     \fi
\ifx \showISSN     \undefined \def \showISSN      #1{\unskip}     \fi
\ifx \showLCCN     \undefined \def \showLCCN      #1{\unskip}     \fi
\ifx \shownote     \undefined \def \shownote      #1{#1}          \fi
\ifx \showarticletitle \undefined \def \showarticletitle #1{#1}   \fi
\ifx \showURL      \undefined \def \showURL       {\relax}        \fi
\providecommand\bibfield[2]{#2}
\providecommand\bibinfo[2]{#2}
\providecommand\natexlab[1]{#1}
\providecommand\showeprint[2][]{arXiv:#2}

\bibitem[Carlini and Wagner(2017)]%
        {carlini2017towards}
\bibfield{author}{\bibinfo{person}{Nicholas Carlini} {and} \bibinfo{person}{David Wagner}.} \bibinfo{year}{2017}\natexlab{}.
\newblock \showarticletitle{Towards evaluating the robustness of neural networks}. In \bibinfo{booktitle}{\emph{2017 ieee symposium on security and privacy (sp)}}. Ieee, \bibinfo{pages}{39--57}.
\newblock


\bibitem[Chakraborty et~al\mbox{.}(2018)]%
        {chakraborty2018adversarial}
\bibfield{author}{\bibinfo{person}{Anirban Chakraborty}, \bibinfo{person}{Manaar Alam}, \bibinfo{person}{Vishal Dey}, \bibinfo{person}{Anupam Chattopadhyay}, {and} \bibinfo{person}{Debdeep Mukhopadhyay}.} \bibinfo{year}{2018}\natexlab{}.
\newblock \showarticletitle{Adversarial attacks and defences: A survey}.
\newblock \bibinfo{journal}{\emph{arXiv preprint arXiv:1810.00069}} (\bibinfo{year}{2018}).
\newblock


\bibitem[Ding et~al\mbox{.}(2023)]%
        {ding2023emshepherd}
\bibfield{author}{\bibinfo{person}{Ruyi Ding}, \bibinfo{person}{Cheng Gongye}, \bibinfo{person}{Siyue Wang}, \bibinfo{person}{A~Adam Ding}, {and} \bibinfo{person}{Yunsi Fei}.} \bibinfo{year}{2023}\natexlab{}.
\newblock \showarticletitle{EMShepherd: Detecting Adversarial Samples via Side-channel Leakage}. In \bibinfo{booktitle}{\emph{ACM Asia Conference on Computer and Communications Security}}.
\newblock


\bibitem[Feinman et~al\mbox{.}(2017)]%
        {feinman2017detecting}
\bibfield{author}{\bibinfo{person}{Reuben Feinman}, \bibinfo{person}{Ryan~R Curtin}, \bibinfo{person}{Saurabh Shintre}, {and} \bibinfo{person}{Andrew~B Gardner}.} \bibinfo{year}{2017}\natexlab{}.
\newblock \showarticletitle{Detecting adversarial samples from artifacts}.
\newblock \bibinfo{journal}{\emph{arXiv:1703.00410}} (\bibinfo{year}{2017}).
\newblock


\bibitem[Fu et~al\mbox{.}(2020)]%
        {fu2020detecting}
\bibfield{author}{\bibinfo{person}{Hao Fu}, \bibinfo{person}{Akshaj~Kumar Veldanda}, \bibinfo{person}{Prashanth Krishnamurthy}, \bibinfo{person}{Siddharth Garg}, {and} \bibinfo{person}{Farshad Khorrami}.} \bibinfo{year}{2020}\natexlab{}.
\newblock \showarticletitle{Detecting backdoors in neural networks using novel feature-based anomaly detection}.
\newblock \bibinfo{journal}{\emph{arXiv preprint arXiv:2011.02526}} (\bibinfo{year}{2020}).
\newblock


\bibitem[Goodfellow et~al\mbox{.}(2014)]%
        {goodfellow2014explaining}
\bibfield{author}{\bibinfo{person}{Ian~J Goodfellow}, \bibinfo{person}{Jonathon Shlens}, {and} \bibinfo{person}{Christian Szegedy}.} \bibinfo{year}{2014}\natexlab{}.
\newblock \showarticletitle{Explaining and harnessing adversarial examples}.
\newblock \bibinfo{journal}{\emph{arXiv preprint arXiv:1412.6572}} (\bibinfo{year}{2014}).
\newblock


\bibitem[Gravellier(2021)]%
        {gravellier2021remote}
\bibfield{author}{\bibinfo{person}{Joseph Gravellier}.} \bibinfo{year}{2021}\natexlab{}.
\newblock \emph{\bibinfo{title}{Remote hardware attacks on connected devices}}.
\newblock \bibinfo{thesistype}{Ph.\,D. Dissertation}. \bibinfo{school}{Ecole des Mines de Saint-Etienne}.
\newblock


\bibitem[Ilyas et~al\mbox{.}(2018)]%
        {ilyas2018black}
\bibfield{author}{\bibinfo{person}{Andrew Ilyas}, \bibinfo{person}{Logan Engstrom}, \bibinfo{person}{Anish Athalye}, {and} \bibinfo{person}{Jessy Lin}.} \bibinfo{year}{2018}\natexlab{}.
\newblock \showarticletitle{Black-box adversarial attacks with limited queries and information}. In \bibinfo{booktitle}{\emph{International Conference on Machine Learning}}.
\newblock


\bibitem[Jain et~al\mbox{.}(2022)]%
        {jain2022adversarial}
\bibfield{author}{\bibinfo{person}{Shubham Jain}, \bibinfo{person}{Ana-Maria Crețu}, {and} \bibinfo{person}{Yves-Alexandre de Montjoye}.} \bibinfo{year}{2022}\natexlab{}.
\newblock \showarticletitle{Adversarial Detection Avoidance Attacks: Evaluating the robustness of perceptual hashing-based client-side scanning}. In \bibinfo{booktitle}{\emph{USENIX Security Symposium}}.
\newblock


\bibitem[Jin et~al\mbox{.}(2020)]%
        {jin2020unified}
\bibfield{author}{\bibinfo{person}{Kaidi Jin}, \bibinfo{person}{Tianwei Zhang}, \bibinfo{person}{Chao Shen}, \bibinfo{person}{Yufei Chen}, \bibinfo{person}{Ming Fan}, \bibinfo{person}{Chenhao Lin}, {and} \bibinfo{person}{Ting Liu}.} \bibinfo{year}{2020}\natexlab{}.
\newblock \showarticletitle{A unified framework for analyzing and detecting malicious examples of dnn models}.
\newblock \bibinfo{journal}{\emph{arXiv preprint arXiv:2006.14871}} \bibinfo{volume}{8}, \bibinfo{number}{9} (\bibinfo{year}{2020}).
\newblock


\bibitem[Juuti et~al\mbox{.}(2019)]%
        {juuti2019prada}
\bibfield{author}{\bibinfo{person}{Mika Juuti}, \bibinfo{person}{Sebastian Szyller}, \bibinfo{person}{Samuel Marchal}, {and} \bibinfo{person}{N Asokan}.} \bibinfo{year}{2019}\natexlab{}.
\newblock \showarticletitle{PRADA: protecting against DNN model stealing attacks}. In \bibinfo{booktitle}{\emph{2019 IEEE European Symposium on Security and Privacy (EuroS\&P)}}. IEEE, \bibinfo{pages}{512--527}.
\newblock


\bibitem[Kariyappa and Qureshi(2020)]%
        {kariyappa2020defending}
\bibfield{author}{\bibinfo{person}{Sanjay Kariyappa} {and} \bibinfo{person}{Moinuddin~K Qureshi}.} \bibinfo{year}{2020}\natexlab{}.
\newblock \showarticletitle{Defending against model stealing attacks with adaptive misinformation}. In \bibinfo{booktitle}{\emph{Proceedings of the IEEE/CVF Conference on Computer Vision and Pattern Recognition}}. \bibinfo{pages}{770--778}.
\newblock


\bibitem[Kwon(2020)]%
        {kwon2020detecting}
\bibfield{author}{\bibinfo{person}{Hyun Kwon}.} \bibinfo{year}{2020}\natexlab{}.
\newblock \showarticletitle{Detecting backdoor attacks via class difference in deep neural networks}.
\newblock \bibinfo{journal}{\emph{IEEE Access}}  \bibinfo{volume}{8} (\bibinfo{year}{2020}), \bibinfo{pages}{191049--191056}.
\newblock


\bibitem[Li et~al\mbox{.}(2022)]%
        {li2022backdoor}
\bibfield{author}{\bibinfo{person}{Yiming Li}, \bibinfo{person}{Yong Jiang}, \bibinfo{person}{Zhifeng Li}, {and} \bibinfo{person}{Shu-Tao Xia}.} \bibinfo{year}{2022}\natexlab{}.
\newblock \showarticletitle{Backdoor learning: A survey}.
\newblock \bibinfo{journal}{\emph{IEEE Transactions on Neural Networks and Learning Systems}} (\bibinfo{year}{2022}).
\newblock


\bibitem[Li et~al\mbox{.}(2021)]%
        {li2021backdoor}
\bibfield{author}{\bibinfo{person}{Yiming Li}, \bibinfo{person}{Tongqing Zhai}, \bibinfo{person}{Yong Jiang}, \bibinfo{person}{Zhifeng Li}, {and} \bibinfo{person}{Shu-Tao Xia}.} \bibinfo{year}{2021}\natexlab{}.
\newblock \showarticletitle{Backdoor attack in the physical world}.
\newblock \bibinfo{journal}{\emph{arXiv preprint arXiv:2104.02361}} (\bibinfo{year}{2021}).
\newblock


\bibitem[Madry et~al\mbox{.}(2017)]%
        {madry2017towards}
\bibfield{author}{\bibinfo{person}{Aleksander Madry}, \bibinfo{person}{Aleksandar Makelov}, \bibinfo{person}{Ludwig Schmidt}, \bibinfo{person}{Dimitris Tsipras}, {and} \bibinfo{person}{Adrian Vladu}.} \bibinfo{year}{2017}\natexlab{}.
\newblock \showarticletitle{Towards deep learning models resistant to adversarial attacks}.
\newblock \bibinfo{journal}{\emph{arXiv preprint arXiv:1706.06083}} (\bibinfo{year}{2017}).
\newblock


\bibitem[Moini et~al\mbox{.}({[n.\,d.]})]%
        {0Understanding}
\bibfield{author}{\bibinfo{person}{S. Moini}, \bibinfo{person}{X. Li}, \bibinfo{person}{P. Stanwicks}, \bibinfo{person}{G. Provelengios}, \bibinfo{person}{W. Burleson}, \bibinfo{person}{R. Tessier}, {and} \bibinfo{person}{D. Holcomb}.} \bibinfo{year}{[n.\,d.]}\natexlab{}.
\newblock \showarticletitle{Understanding and Comparing the Capabilities of On-Chip Voltage Sensors against Remote Power Attacks on FPGAs}. In \bibinfo{booktitle}{\emph{2020 IEEE 63rd International Midwest Symposium on Circuits and Systems (MWSCAS)}}.
\newblock


\bibitem[Moosavi-Dezfooli et~al\mbox{.}(2016)]%
        {moosavi2016deepfool}
\bibfield{author}{\bibinfo{person}{Seyed-Mohsen Moosavi-Dezfooli}, \bibinfo{person}{Alhussein Fawzi}, {and} \bibinfo{person}{Pascal Frossard}.} \bibinfo{year}{2016}\natexlab{}.
\newblock \showarticletitle{Deepfool: a simple and accurate method to fool deep neural networks}. In \bibinfo{booktitle}{\emph{Proceedings of the IEEE conference on computer vision and pattern recognition}}. \bibinfo{pages}{2574--2582}.
\newblock


\bibitem[Odetola et~al\mbox{.}(2022)]%
        {odetola2022hardening}
\bibfield{author}{\bibinfo{person}{Tolulope~A Odetola}, \bibinfo{person}{Adewale Adeyemo}, {and} \bibinfo{person}{Syed~Rafay Hasan}.} \bibinfo{year}{2022}\natexlab{}.
\newblock \showarticletitle{Hardening hardware accelerartor based CNN inference phase against adversarial noises}. In \bibinfo{booktitle}{\emph{IEEE International Symposium on Hardware Oriented Security and Trust}}.
\newblock


\bibitem[Papernot et~al\mbox{.}(2016)]%
        {papernot2016practical}
\bibfield{author}{\bibinfo{person}{Nicolas Papernot}, \bibinfo{person}{Patrick McDaniel}, \bibinfo{person}{Ian Goodfellow}, \bibinfo{person}{Somesh Jha}, \bibinfo{person}{Z~Berkay Celik}, {and} \bibinfo{person}{Ananthram Swami}.} \bibinfo{year}{2016}\natexlab{}.
\newblock \showarticletitle{Practical black-box attacks against deep learning systems using adversarial examples}.
\newblock \bibinfo{journal}{\emph{arXiv preprint arXiv:1602.02697}} (\bibinfo{year}{2016}).
\newblock


\bibitem[Rigaki and Garcia(2023)]%
        {rigaki2023survey}
\bibfield{author}{\bibinfo{person}{Maria Rigaki} {and} \bibinfo{person}{Sebastian Garcia}.} \bibinfo{year}{2023}\natexlab{}.
\newblock \showarticletitle{A survey of privacy attacks in machine learning}.
\newblock \bibinfo{journal}{\emph{Comput. Surveys}} \bibinfo{volume}{56}, \bibinfo{number}{4} (\bibinfo{year}{2023}), \bibinfo{pages}{1--34}.
\newblock


\bibitem[Salimans et~al\mbox{.}(2017)]%
        {salimans2017evolution}
\bibfield{author}{\bibinfo{person}{Tim Salimans}, \bibinfo{person}{Jonathan Ho}, \bibinfo{person}{Xi Chen}, \bibinfo{person}{Szymon Sidor}, {and} \bibinfo{person}{Ilya Sutskever}.} \bibinfo{year}{2017}\natexlab{}.
\newblock \showarticletitle{Evolution strategies as a scalable alternative to reinforcement learning}.
\newblock \bibinfo{journal}{\emph{arXiv preprint arXiv:1703.03864}} (\bibinfo{year}{2017}).
\newblock


\bibitem[Samavatian et~al\mbox{.}(2021)]%
        {samavatian2021hasi}
\bibfield{author}{\bibinfo{person}{Mohammad~Hossein Samavatian}, \bibinfo{person}{Saikat Majumdar}, \bibinfo{person}{Kristin Barber}, {and} \bibinfo{person}{Radu Teodorescu}.} \bibinfo{year}{2021}\natexlab{}.
\newblock \showarticletitle{HASI: Hardware-accelerated stochastic inference, a defense against adversarial machine learning attacks}.
\newblock \bibinfo{journal}{\emph{arXiv:2106.05825}} (\bibinfo{year}{2021}).
\newblock


\bibitem[Selvaraju et~al\mbox{.}(2016)]%
        {selvaraju2016grad}
\bibfield{author}{\bibinfo{person}{Ramprasaath~R Selvaraju}, \bibinfo{person}{Abhishek Das}, \bibinfo{person}{Ramakrishna Vedantam}, \bibinfo{person}{Michael Cogswell}, \bibinfo{person}{Devi Parikh}, {and} \bibinfo{person}{Dhruv Batra}.} \bibinfo{year}{2016}\natexlab{}.
\newblock \showarticletitle{Grad-CAM: Why did you say that?}
\newblock \bibinfo{journal}{\emph{arXiv preprint arXiv:1611.07450}} (\bibinfo{year}{2016}).
\newblock


\bibitem[Xu et~al\mbox{.}(2017)]%
        {xu2017feature}
\bibfield{author}{\bibinfo{person}{Weilin Xu}, \bibinfo{person}{David Evans}, {and} \bibinfo{person}{Yanjun Qi}.} \bibinfo{year}{2017}\natexlab{}.
\newblock \showarticletitle{Feature squeezing: Detecting adversarial examples in deep neural networks}.
\newblock \bibinfo{journal}{\emph{arXiv:1704.01155}} (\bibinfo{year}{2017}).
\newblock


\bibitem[Xu et~al\mbox{.}(2021)]%
        {xu2021detecting}
\bibfield{author}{\bibinfo{person}{Xiaojun Xu}, \bibinfo{person}{Qi Wang}, \bibinfo{person}{Huichen Li}, \bibinfo{person}{Nikita Borisov}, \bibinfo{person}{Carl~A Gunter}, {and} \bibinfo{person}{Bo Li}.} \bibinfo{year}{2021}\natexlab{}.
\newblock \showarticletitle{Detecting ai trojans using meta neural analysis}. In \bibinfo{booktitle}{\emph{2021 IEEE Symposium on Security and Privacy (SP)}}. IEEE, \bibinfo{pages}{103--120}.
\newblock


\bibitem[Yan et~al\mbox{.}(2023)]%
        {yan2023mercury}
\bibfield{author}{\bibinfo{person}{Xiaobei Yan}, \bibinfo{person}{Xiaoxuan Lou}, \bibinfo{person}{Guowen Xu}, \bibinfo{person}{Han Qiu}, \bibinfo{person}{Shangwei Guo}, \bibinfo{person}{Chip~Hong Chang}, {and} \bibinfo{person}{Tianwei Zhang}.} \bibinfo{year}{2023}\natexlab{}.
\newblock \showarticletitle{Mercury: An Automated Remote Side-channel Attack to Nvidia Deep Learning Accelerator}. In \bibinfo{booktitle}{\emph{IEEE International Conference on Field-Programming Technology}}.
\newblock


\bibitem[Zhao and Suh(2018)]%
        {zhao2018fpga}
\bibfield{author}{\bibinfo{person}{Mark Zhao} {and} \bibinfo{person}{G~Edward Suh}.} \bibinfo{year}{2018}\natexlab{}.
\newblock \showarticletitle{FPGA-based remote power side-channel attacks}. In \bibinfo{booktitle}{\emph{IEEE Symposium on Security and Privacy}}.
\newblock


\end{thebibliography}

\end{document}